\documentclass[iop,apj]{emulateapj} 
\usepackage{color}

\usepackage{pifont} 
\shorttitle{Galilean Satellites Brightness in Eclipse}
\shortauthors{Tsumura et al.}

\begin{document}

\title{Near-infrared Brightness of the Galilean Satellites Eclipsed in Jovian Shadow:\\ A New Technique to Investigate Jovian Upper Atmosphere}
\author{K. Tsumura\altaffilmark{1}, K. Arimatsu\altaffilmark{2,3}, E. Egami\altaffilmark{4}, Y. Hayano\altaffilmark{5}, C. Honda\altaffilmark{6}, J. Kimura\altaffilmark{7}, K. Kuramoto\altaffilmark{8}, S. Matsuura\altaffilmark{2},\\ Y. Minowa\altaffilmark{5}, K. Nakajima\altaffilmark{9}, T. Nakamoto\altaffilmark{10}, M. Shirahata\altaffilmark{11,2}, J. Surace\altaffilmark{12}, Y. Takahashi\altaffilmark{8}, and T. Wada\altaffilmark{2}}
 
\altaffiltext{1}{Frontier Research Institute for Interdisciplinary Science, Tohoku University, Sendai, Miyagi 980-8578, Japan}     
\altaffiltext{2}{Department of Space Astronomy and Astrophysics, Institute of Space and Astronoutical Science, Japan Aerospace Exploration Agency, Sagamihara, Kanagawa 252-5210, Japan}
\altaffiltext{3}{Department of Astronomy, Graduate School of Science, The University of Tokyo, Tokyo 113-0033, Japan}
\altaffiltext{4}{Department of Astronomy, Arizona University, Tucson, AZ 85721, USA}
\altaffiltext{5}{Hawaii Observatory, National Astronomical Observatory of Japan, Hilo, HI 96720, USA}
\altaffiltext{6}{Research Center for Advanced Information Science and Technology, Aizu Research Cluster for Space Science, The University of Aizu, Aizu-Wakamatsu, Fukushima 965-8589, Japan}
\altaffiltext{7}{Earth-Life Science Institute, Tokyo Institute of Technology, Tokyo 152-8550, Japan}
\altaffiltext{8}{Department of Cosmosciences, Graduate School of Science, Hokkaido University, Sapporo, Hokkaido 060-0810, Japan}
\altaffiltext{9}{Department of Earth and Planetary Sciences, Kyushu University, Fukuoka 812-8581, Japan}
\altaffiltext{10}{Department of Earth and Planetary Sciences, Graduate School of Science and Engineering, Tokyo Institute of Technology, Tokyo 152-8551, Japan}
\altaffiltext{11}{National Astronomical Observatory of Japan, Mitaka, Tokyo 181-8588, Japan}
\altaffiltext{12}{Spitzer Science Center, California Institute of Technology, Pasadena, CA 91125, USA}
\email{tsumura@astr.tohoku.ac.jp}

\begin{abstract}
We have discovered that Europa, Ganymede and Callisto are bright around 1.5 $\mu$m even when not directly lit by sunlight, 
based on observations from the Hubble Space Telescope and the Subaru Telescope.  
The observations were conducted with non-sidereal tracking on Jupiter outside of the field of view to reduce the stray light subtraction uncertainty due to the close proximity of Jupiter.
Their eclipsed luminosity was 10$^{-6}$--10$^{-7}$ of their uneclipsed brightness, which is low enough that this phenomenon has been undiscovered until now.  
In addition, Europa in eclipse was $<$1/10 of the others at 1.5 $\mu$m, a potential clue to the origin of the source of luminosity.  
Likewise, Ganymede observations were attempted at 3.6 $\mu$m by the Spitzer Space Telescope but it was not detected, suggesting a significant wavelength dependence.  
The reason why they are luminous even when in the Jovian shadow is still unknown, but forward-scattered sunlight by haze in the Jovian upper atmosphere is proposed as the most plausible candidate. 
If this is the case, observations of these Galilean satellites while eclipsed by the Jovian shadow provide us a new technique to investigate Jovian atmospheric composition,
and investigating the transmission spectrum of Jupiter by this method is important for investigating the atmosphere of extrasolar giant planets by transit spectroscopy.
\end{abstract}
\keywords{atmospheric effects --- eclipses --- planets and satellites: individual (Jupiter, Europa, Ganymede, Callisto)}

\section{Introduction}
The Galilean satellites (Io, Europa, Ganymede, and Callisto) around Jupiter undergo frequent eclipses in the Jovian shadow.
Because their brightness out of eclipse is dominated by sunlight illumination and thermal emission from their $\sim$120 K surface is negligible in the optical and near-infrared wavelengths,
the Galilean satellites in eclipse are expected to be dark in optical and near-infrared wavelengths except Io, 
whose thermal emission from volcanoes can be observed during its eclipse \citep{Pater2004}.  
These eclipse events provide us unique opportunities for a variety of studies.  
The historic first measurement of the speed of light was achieved by observations of eclipses of Galilean satellites \citep{Romer}, 
and recently this kind of observation has been used to obtain the accurate astrometric data with a great precision \citep{Emelyanov2006, Emelyanov2009, Mallama2010, Emelyanov2011}.  
Jovian upper atmosphere was investigated by the observations of eclipse events of Galilean satellites \citep{Smith1977, Greene1980, Smith1980-1, Smith1980-2}, 
which allow us to study the Jovian atmosphere by the astronomical observations from the Earth, not by the in-situ probes of spacecraft missions to Jupiter.  
These studies using eclipses of Galilean satellites are based on the observations of light curve of ingress or egress of eclipses \citep{Mallama1991, Mallama1992}, 
because nothing is expected to be detected by observations of these satellites in eclipse.

\begin{table*}
\begin{center}
\caption{Our observations of eclipses \label{table}}
\begin{tabular}{ccccccc}
\tableline\tableline
Satellite & Date (UT) & Telescope & Instrument & Filter & Impact parameter & Brightness\\
\tableline
Europa & 2012/Feb./21 & Subaru & IRCS & J-band (1.25 $\mu$m) & 0.57-0.62 & $<$1.5 $\mu$Jy \\
Europa & 2013/Apr./8 & Hubble & WFC3 & F139M (1.39 $\mu$m) & 0.54-0.76 & $<$5.5 $\mu$Jy \\
Europa & 2013/Nov./18 & Subaru & IRCS & CH$_{4}$-long (1.69 $\mu$m) & 0.76-0.85 & $<$88 $\mu$Jy \tablenotemark{a} \\
Europa & 2014/Mar./19 & Subaru & IRCS& CH$_{4}$-long (1.69 $\mu$m)  & 0.32-0.96 & $<$0.21 $\mu$Jy  \\
Europa & 2014/Mar./26 & Hubble & WFC3 & F139M (1.39 $\mu$m) & 0.30-0.65 & 6.0-9.5 $\mu$Jy \\
Ganymede & 2012/Mar./26 & Spitzer & IRAC & Channel 1 (3.6 $\mu$m) & $>$0.87 & $<$3.6 $\mu$Jy \\
Ganymede & 2012/Jul./26 & Subaru & IRCS & J-band (1.25 $\mu$m) & 0.86-0.95 & 60-100 $\mu$Jy \\
Ganymede & 2013/Feb./5 & Hubble & WFC3 & F160W (1.60 $\mu$m) & 0.77-0.94 & 60-80 $\mu$Jy \\
Ganymede & 2013/Mar./5 & Hubble & WFC3 & F139M (1.39 $\mu$m) & 0.79-0.74 & 25-35 $\mu$Jy \\
Callisto & 2013/Oct./20 & Subaru & IRCS & J-band (1.25 $\mu$m) & 0.88-0.94 & 20-40 $\mu$Jy \\
\tableline
& & & & & & $^a$ Bad weather 
\end{tabular}
\end{center}
\vskip -0.4cm
\end{table*}

Here, we report unexpected detections of Europa, Ganymede, and Callisto eclipsed in Jovian shadow at around 1.5 $\mu$m using the Hubble Space Telescope and the Subaru Telescope.
Our motivation of observing the Galilean satellites in eclipse by such great telescopes is 
to detect the extragalactic background light in the near-infrared wavelengths \citep{Cam01, Matsumoto05, Tsumura2013} 
without any zodiacal light subtraction uncertainty by using Galilean satellites eclipsed by Jupiter as occulting spots.
Thus we expected that the Galilean satellites in eclipse are dark enough as occulters in the near-infrared wavelengths, but we found they are bright even in the Jovian shadow.
Light source of the brightness of these satellites in eclipse remain inconclusive although some candidates are discussed in this paper, 
and that is why this new finding has a potential to bring us new insights about Jupiter and/or Galilean satellites.
For example, forward-scattered sunlight by haze in the Jovian upper atmosphere is suggested as the most probable candidate of illuminator in this paper and \citet{Nakamoto2014}.  
If this is the case, by monitoring many eclipses with different configurations, we can investigate the composition of the Jovian atmosphere, especially abundance of methane and haze particles,
as a function of position (latitude and altitude) of the atmosphere by observations from the Earth.  
We think that this method makes it possible to estimate the information of haze particles around the top of the stratosphere of Jovian atmosphere, 
where the cloud and haze are mainly produced by photochemistry and play important role in the cloud formation as condensation nuclei, but poorly known.
Since their brightness in eclipse was 10$^{-6}$--10$^{-7}$ of their brightness out of eclipse, 
this phenomenon has been undiscovered until now and the state-of-the-art telescopes are required to detected such dark signals in limited duration time of eclipses.

\section{Method}
\subsection{Data Acquisition}
Our observations were conducted with the Infrared Camera and Spectrograph (IRCS) \citep{Kobayashi2000} using the adaptive optics (AO188) \citep{Hayano2010} with natural guide star mode on the Subaru telescope \citep{Iye2004}, 
the Wide Field Camera 3 (WFC3) \citep{MacKenty2012} on the Hubble Space Telescope, and the Infrared Array Camera (IRAC) \citep{Fazio2004} on the Spitzer Space Telescope \citep{Werner2004}.  
Our observations are summarized in Table \ref{table}.
Since our observational targets are specific astronomical events, chances for observation are limited and timing is absolutely critical.
The separation from the Jovian limb to the satellites is less than 1 arcmin in our observations, thus the greatest difficulty results from stray light due to the close proximity of Jupiter to the target satellites. 
Therefore we developed an advanced method for our observations. 
Although details of our method depend on each observation, a basic strategy is as follows: 
Jupiter was kept out of the detector field of view (FoV) during the observation to avoid electric crosstalk on the detector array and stray light from Jupiter, 
which is $\sim10^8$ brighter than the eclipsed satellites.  
Observations were conducted with non-sidereal tracking on Jupiter (outside of the FoV) to fix the stray light pattern on the detector during the observation\footnote{Only the observation of Ganymede eclipse by Hubble in 2013/Feb./5 was conducted with tracking on Ganymede.}.  
This minimizes the systematic error due to the Jovian stray light.
Owing to absorption by methane in Jovian atmosphere, Jupiter is dark in methane bands (CH$_4$-long filter for Subaru/IRCS, F139M for Hubble/WFC3, and Channel 1 for Spitzer/IRAC),
so observations in the methane bands are very effective to reduce the stray light from Jupiter.
Short integration times were required to avoid smearing of the target satellite owing to the relative movement of the target satellites to Jupiter.  
Letting the eclipsed satellite move relative to the detector using Jupiter tracking has the advantage of effectively dithering the observations so that we can average out any detector issues.  
The satellite ephemerides are provided with great precision by the Natural Satellites Ephemeride Server MULTI-SAT\footnote{http://www.sai.msu.ru/neb/nss/nssphe0he.htm} \citep{EmelYanov2008}
and the JPL-HORIZONS\footnote{http://ssd.jpl.nasa.gov/horizons.cgi} \citep{Giorgini1996},
the size and location of the eclipsed satellite are also accurately known, and therefore easily differentiated from detector effects.

\begin{figure*}[htbp]
\begin{center}
\includegraphics[scale=0.52]{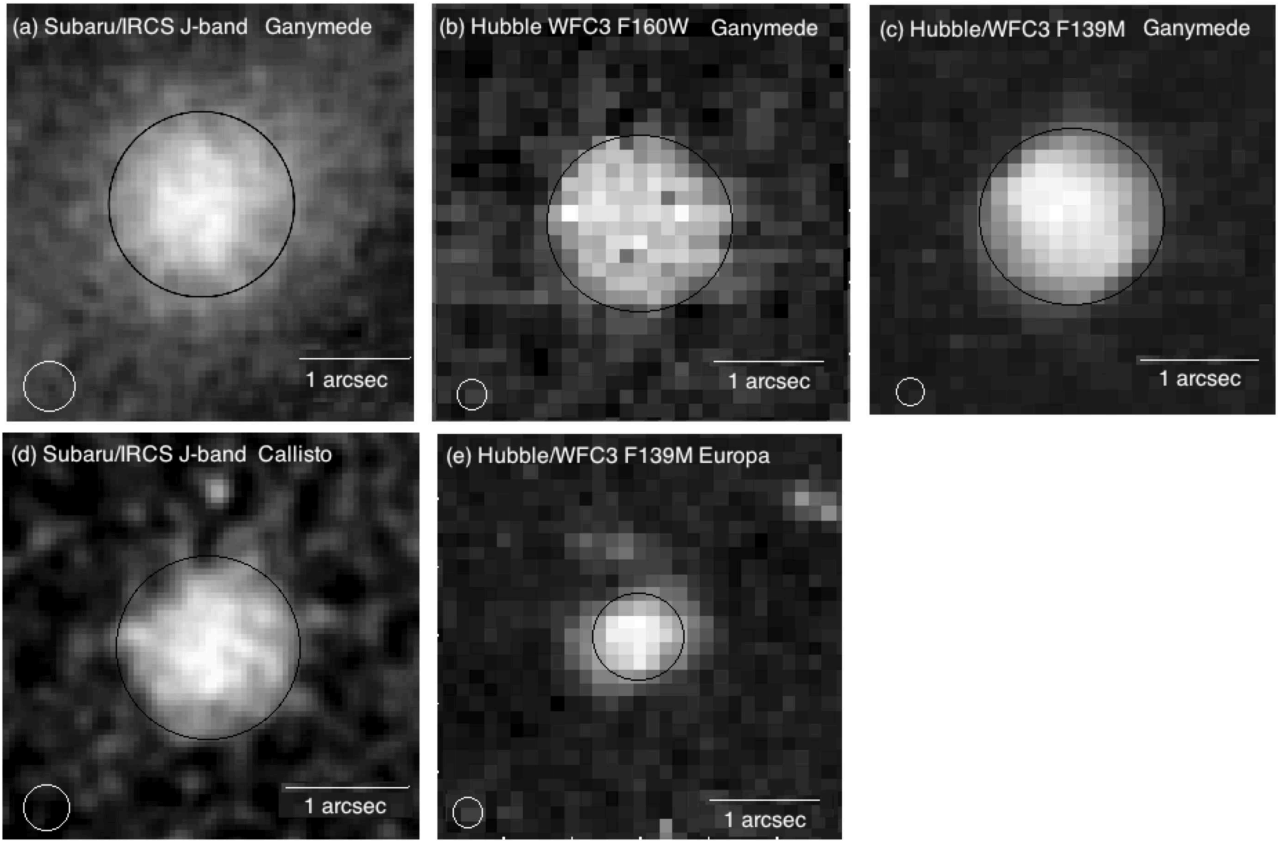}
\end{center}
\vskip -0.5cm
\caption{Images of the Galilean satellites eclipsed by Jovian shadow.
               Image (a) was Ganymede obtained by Subaru/IRCS with J-band filter in 2012/Jul./26 (UT), 
               image (b) was Ganymede obtained by Hubble/WFC3 with F160W filter in 2013/Feb./5 (UT), 
               image (c) was Ganymede obtained by Hubble/WFC3 with F139M filter in 2013/Mar./5 (UT), 
               image (d) was Callisto obtained by Subaru/IRCS with J-band filter in 2013/Oct./20 (UT),
               and image (e) was Europa obtained by Hubble/WFC3 with F139M filter in 2014/Mar./26 (UT)  
               The Subaru images are convolved with a Gausian with FWHM of 2 pixels.
               The centre black circles show the size of Ganymede or Callisto, and the small white circles at the bottom left corner show the PSF size (FWHM) in each image.\label{image}}
\end{figure*}

\subsection{Data Processing}
In addition to the standard data reduction such as dark frame subtraction, flat-field correction, and calibration using standard stars,
we conducted a special data processing described below.

In each observation we obtained a time series of the observed images, in which the target satellite in eclipse was moving in the fixed stray light pattern from Jupiter outside of FoV.
The stray light pattern for each single frame was constructed from other images in which the target satellites moved more than the satellite size (0.8--1.6 arcsec) from the single frame, and was subtracted.
In the case of observations with the methane bands, the stray light level from Jupiter is equal to or less than the sky level, 
so the uncertainty due to the stray light subtraction is also equal to or less than the uncertainty from photon noise from the sky level.
In the case of observations with non-methane bands (J-band filter for Subaru/IRCS and F160W for Hubble/WFC3), the stray light level from Jupiter was five to twelve times brighter than the sky level,
so the uncertainty from photon noise from the stray light was added to that from the sky level.
However, the stray light pattern from Jupiter was successfully canceled out thanks to the observations with Jupiter tracking, thus systematic uncertainty from the stray light subtraction should be negligible.
Another Galilean satellite out of eclipse ($\sim10^6$ brighter than the eclipsed satellites) was in the images in some data 
(Subaru/IRCS data in 2012/Feb./21 and 2012/Jul./26, and Hubble/WFC3 data in 2013/Feb./5 and 2013/Apr./8),
and the target satellite in eclipse was contaminated by the near-by bright satellite.
Since the satellites of the contamination source had a different movement relative to the target satellite in eclipse and Jupiter, contamination pattern was not fixed in the images.
Even in such a case, since the contamination pattern does not have a local structure, the contamination pattern was evaluated around the target satellite and subtracted locally\footnote{Since the observation of Ganymede eclipse by Hubble in 2013/Feb./5 was conducted with tracking on Ganymede, the stray light subtraction using the fixed pattern was not applied. 
However, since Europa out of eclipse was located by Ganymede in eclipse, uncertainty by the contamination subtraction from Europa was dominated in this data.}.

After the stray light subtraction, brightness of the target satellite in eclipse was evaluated.
Since we know the exact position and diameter of the satellites, 
and we can correct the positioning information in FITS header of each images by comparing it to field stars and the satellite out of eclipse in the images,
we can find the pixels where the target satellite in eclipse should exist even in the non-detection case of it.
The amounts of positioning correction by comparing to field stars were less than five pixels.
Total uncertainty after the subtraction of the stray light from Jupiter and the satellite out of eclipse was evaluated from the deviation of the pixels around the target satellite in eclipse.

\section{Result}
The brightness of Ganymede in eclipse was $\sim$80 $\mu$Jy ($\sim$19.1 AB mag) at J-band which was $\sim$4$\times10^{-6}$ extinction from the brightness of Ganymede out of eclipse, 
and the brightness of Callisto in eclipse was $\sim$30 $\mu$Jy ($\sim$20.2 AB mag) at J-band which was $\sim$2$\times10^{-6}$ extinction from the brightness of Callisto out of eclipse. 
On the other hand, Ganymede in eclipse was not detected at 3.6 $\mu$m by Spitzer/IRAC, obtaining an upper limit of its brightness as $<$3.6 $\mu$Jy ($>$22.5 AB mag).  
Europa in eclipse was also observed around 1.5 $\mu$m, and the brightness of Europa was much darker than Ganymede and Callisto.
Europa in eclipse was not detected by observations until 2014/Mar./19 obtaining an upper limit of $<$5.5 $\mu$Jy ($>$22.0 AB mag) with F139M filter 
and $<$0.21 $\mu$Jy ($>$25.6 AB mag) with CH$_4$-long filter,
but it was detected in 2014/Mar./26 obtaining $\sim$7 $\mu$Jy ($\sim$21.8 AB mag) with the same F139M filter.
This discrepancy is discussed in Section \ref{reflection}.
Note that non-detection of Europa in eclipse at $<$820 nm by Hubble/WFC2 was also reported \citep{Sparks2010}. 

\begin{figure}[htbp]
\begin{center}
\includegraphics[scale=0.43]{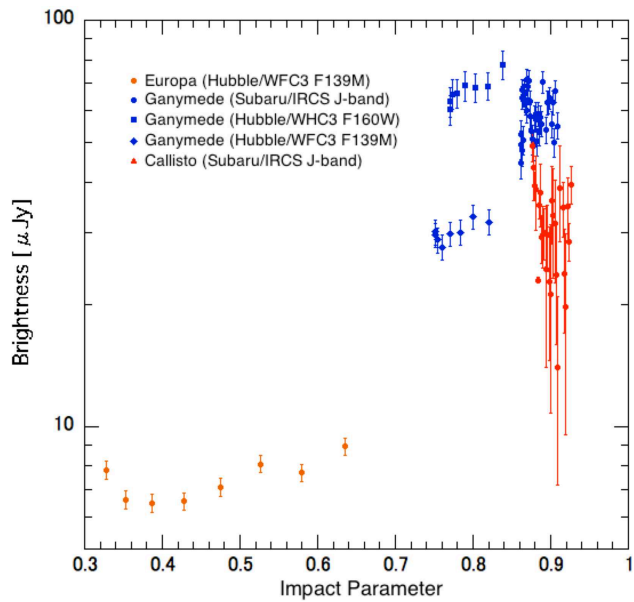}
\end{center}
\vskip -0.2cm
\caption{Brightness variance of observed eclipses as a function of impact parameter. 
Orange points indicate Europa obtained, blue points indicate Ganymede (different symbols show different observations), and red points indicate Callisto.\label{var}}
\vskip 0.1cm
\end{figure}

\begin{figure}[htbp]
\begin{center}
\includegraphics[scale=0.43]{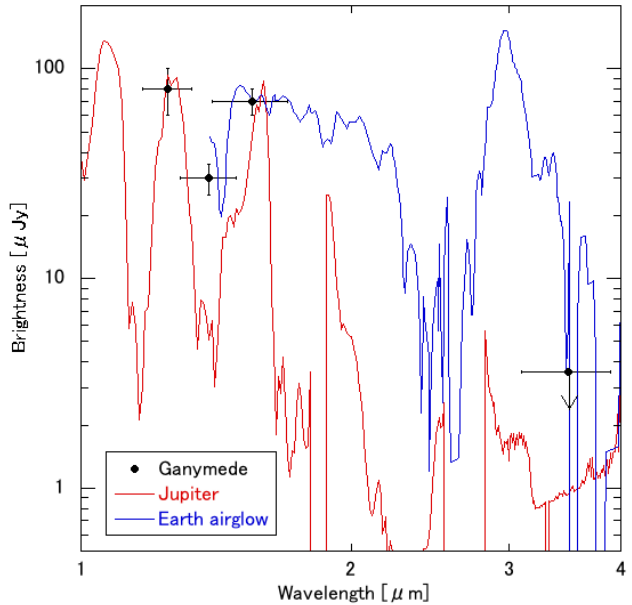}
\end{center}
\vskip -0.6cm
\caption{Spectral energy distribution (SED) of Ganymede in eclipse.  
               Each data point of Ganymede in eclipse was obtained in the different date and impact parameter, 
               thus the geometric relation among Jupiter, Ganymede and the observer is different.  
               Spectra of Jupiter \citep{Rayner2009} and Earth atmosphere \citep{Star1985} are also shown as a reference, scaled to the Ganymede brightness.\label{spec}}
\end{figure}

Figure \ref{image} shows the images of these detected satellites in eclipse.
Since the satellites moved in the images in the time series data, images in Figure \ref{image} were obtained by shift-and-add of the time series images.
Surrounding sky of the target satellites in Figure \ref{image} has no structure, which shows that the stray light was successfully subtracted.
Thanks to the fine spatial resolutions of Hubble/WFC3 and Subaru/IRCS+AO188, images of these satellites were spread over several pixels as shown in Figure \ref{image}, 
and we found that they were uniformly bright in eclipse.  
Figure \ref{var} shows the variance of brightness of the satellites in eclipse as a function of impact parameter\footnote{Impact parameter is defined as the distance of the satellite from the center of the shadow, ratioed to Jovian radius along the target satellite.} from the time series data,  
and small dependences on impact parameter were detected within the shallow eclipse range (impact parameter $>$0.7). 
Subaru data had larger dispersion than Hubble data because of instability of the Earth's atmosphere. 
Figure \ref{spec} shows the broad-band spectral energy distribution (SED) of Ganymede in eclipse.  
Note that relative relation among these data points in Figure \ref{spec} should not be accepted at face value because each data point was obtained in different days, 
thus the geometric relation among Jupiter, Ganymede, and the observer is different.

\section{Discussion}
Why are the Galilean satellites bright in the Jovian shadow?  
We consider the followings as possible explanations: 
(A) atmospheric emission from the satellites, (B) illumination by emission from the dark side of Jupiter, 
(C) illumination by sunlight reflected from the other satellites,
and (D) illumination by forward-scattered sunlight at the Jovian upper atmosphere to the satellites.

\subsection{Atmospheric emission from the satellites}
One possible explanation is (A) the atmospheric emission from the satellites in eclipse.  
Auroral emission from electron excited atomic oxygen at UV (130.4 and 135.6 nm) \citep{Hall1998, McGrath2013} and optical (630.0 and 636.3 nm) \citep{Brown1999} at both poles of Ganymede in eclipse was known, 
but this auroral emission cannot explain the uniformity of brightness as shown in Figure \ref{image}.  
Thus the origin of this brightness should be radiation from whole atmosphere of the satellites.  
In the case of the Earth atmosphere, OH molecules are the main carrier of the airglow emission in the near-infrared wavelengths, 
which are excited by sunlight at dayside and emit photons at nightside.  
The Galilean satellites are also expected to have OH molecules in their atmosphere, 
since OH molecules can be generated from water ice on their surfaces by sputtering owing to interaction with the plasma particles around Jupiter.  
Such OH molecules in their atmosphere can be excited by the sunlight during out of eclipse and emit photons during their eclipses. 
The column density of OH is estimated to be 10$^{10}$--10$^{11}$ cm$^{-2}$ in Ganymede atmosphere \citep{Marconi2007}, 
which is comparable to that of the OH layer in the Earth atmosphere.  
However, Ganymede SED, especially the non-detection at 3.6 $\mu$m, is difficult to explain by OH airglow because there is an emission band ($\Delta\nu$=1) at 2.5--3.5 $\mu$m \citep{Star1985} as shown in Figure \ref{spec}, 
although excitation state of OH molecules and chemical environment in the atmosphere of the Earth and the Galilean satellites are different, 
resulting in different spectral shape.  
Europa has similar atmosphere to Ganymede and Callisto in density and composition, thus it is also difficult to explain the big difference of brightness between Europa and Ganymede/Callisto.
In fact, atmospheric emissions were detected from Europa but none were detected from Ganymede by the Cassini spacecraft during fly-by \citep{Porco2003},
which is not consistent with this explanation.

\subsection{Emission from the dark side of Jupite}
The second candidate is (B) illumination by emission from the dark side of Jupiter.  
Lightning \citep{Dyudina2004}, aurora \citep{Kim1993, Gladstone2007, Radioti2011}, and nightside airglow \citep{Gladstone2007} of Jupiter are known and they can be candidates of the illuminator.  
However, it is difficult for these illuminators to explain the fact that the brightness of Europa is much darker than Ganymede and Callisto because Europa is located more closely to Jupiter,
thus Europa should be illuminated more strongly from Jupiter than the others.
In addition, the brightness of the satellites in eclipse was kept during our observations (more than one hour in some observations), 
thus it is also difficult to be explained by sporadic events like lightning and aurora.

\subsection{Sunlight reflected from the other satellites}\label{reflection}
We estimated the effect of (C) illumination by the reflected sunlight from the other satellites.
Assuming that Io out of eclipse as an illuminator is located $3 \times 10^5$ km away from Europa in eclipse\footnote{Orbital radii from Jupiter are $4.2 \times 10^5$ km for Io, $6.7 \times 10^5$ km for Europa, 
$1.1 \times 10^6$ km for Ganymede, and $1.9 \times 10^6$ km of Callisto.
Thus the assumed distance of $3 \times 10^5$ km is almost the nearest case.},
brightness of Io seen from Europa (with phase angle of 0 degree) is $\sim$6$\times 10^6$ of that seen from the Earth. 
Since extinction owing to phase angle (Sun-Io-Europa) is $>$10$^2$ when the phase angle is $>$135 degrees\footnote{The phase angle (Sun-Io-Europa) of 135 degrees is also the smallest case 
because we consider the situation of Europa in the Jovian shadow (i.e. the angle of Sun-Jupiter-Europa is $\sim$180 degrees).}\citep{Simonelli1984},
brightness of Io seen from Europa with the phase angle of 135 degrees becomes $\sim$6 $\times 10^4$ of that seen from the Earth.
In addition, brightness of the Sun seen from Jupiter (or Io/Europa) is $\sim$4$\times 10^{-2}$ of that seen from the Earth.
We know that the brightness ratio between the Sun and Io seen from the Earth is $\sim$1$\times 10^{13}$,
then we obtained the brightness ratio of them seen from Europa is $<$10$^7$.

This effect cannot explain the detected brightness of Ganymede and Callisto whose brightness was $\sim$10$^{-6}$ of that out of eclipse,
but it can explain the detected brightness of Europa whose brightness was $\sim$10$^{-7}$ of that out of eclipse.
In fact, when we detected the Europa brightness in eclipse in 2014/Mar./26, the phase angle was 134 degrees and the distance between Europa and Io was $3.4 \times 10^5$ km.
These values are almost same as the assumed values in the estimation, meaning that Europa in eclipse was illuminated from Io in the most efficient configuration in this day.
On the other hand, Europa in eclipse was not detected with the same filter in 2013/Apr./8.
In this day, the distance between Europa and Io was $2.8 \times 10^5$ km and the phase angle was 155 degrees.
Io appears brighter by a factor of $\sim$1.5 by the distance effect, but darker by a factor of $>$6 by the phase angle effect \citep{Simonelli1984} than that in 2014/Mar./26.
In the same way, non-detections of Europa in the other days can be explained by the configuration of the Jovian system.
Thus illumination by the reflected sunlight from the other satellites explains both the brightness of Europa in 2014/Mar./26 and the non-detection in the other days.

\subsection{Forward-scattered sunlight at Jovian upper atmosphere}
Thus, we propose a hypothesis that (D) the Galilean satellites are illuminated by sunlight forward-scattered by haze in the Jovian upper atmosphere to explain the brightness of Ganymede and Callisto in eclipse.  
Detailed discussion for this model is found in \citet{Nakamoto2014}, describing that illumination is dominated by haze around the top of the stratosphere of Jovian atmosphere.
Methane absorption-like feature at 1.4 $\mu$m is found in Ganymede SED in eclipse, like in Jovian spectrum \citep{Rayner2009} as shown in Figure \ref{spec}.  
Methane abundance around the top of the stratosphere is almost same as that at higher pressure regions in Jovian atmosphere ($\sim$10$^{-3}$ in mole fraction) \citep{Moses2005}, 
which can explain the 1.4 $\mu$m methane-like absorption feature in Ganymede SED in eclipse if this model is true.  
Albedo of Ganymede at $>$3 $\mu$m is much smaller than that around 1.5 $\mu$m owing to water ice on its surface \citep{Calvin1995}, 
which also explains the non-detection of Ganymede in eclipse at 3.6 $\mu$m.  
In addition, the fact that Europa in eclipse was much darker than the others can be explained by this model 
because the effect of the scattered sunlight should be less at a satellite closer to Jupiter, 
and the depth of the observed Europa eclipses were deeper (impact parameters were smaller) than those of Ganymede and Callisto.  
In such a case, however, extinction of Callisto ($\sim2\times10^{-6}$) should be greater than Ganymede ($\sim4\times10^{-6}$) because Callisto is located farther from Jupiter than Ganymede, but it was not true.  
Difference of albedo and diameter are canceled in this extinction comparison.  
This might be explained by the difference of haze abundance in Jovian upper atmosphere 
because brightness of satellites in eclipse should depend on where (longitude and latitude) the sunlight was scattered in Jovian atmosphere.
Haze abundance in Jovian atmosphere depends on position \citep{Zhang2013}, 
and Callisto might be illuminated by sunlight through the haze rich region of Jovian upper atmosphere when we observed it. 
As shown in Figure \ref{var}, all Ganymede and Callisto eclipses we observed were shallow (impact parameter $>$ 0.7), 
thus they may become darker at deeper eclipses in this hypothesis, which will be tested by our future observations. 
 
As the transmission spectrum of the Earth atmosphere was measured from lunar eclipse observation \citep{Palle2009, Vidal-Madjar2010, Garcia2012}, 
spectra of the Galilean satellites in eclipse could be tracers of Jovian upper atmosphere, 
especially abundance of haze and/or methane along Jovian altitude, if this model is true.  
Particularly, secular change and location dependence of the Jovian atmosphere can be monitored by long-term observations of a number of eclipses.  
For example, observed brightness variances of Ganymede and Callisto shown in Figure \ref{var} might be explained by the variance of haze abundance in Jovian atmosphere.  
Similar studies were conducted to investigate the composition of Jovian atmosphere by Galilean satellites eclipses \citep{Smith1977, Greene1980, Smith1980-1, Smith1980-2}, 
but these previous works are based on observations of light curves of ingress and egress of eclipses illuminated by transmitted 
and refracted sunlight through the deeper part of Jovian upper atmosphere 
($\sim$several hundreds mbar region where is around the bottom of the stratosphere of Jovian atmosphere), 
which cannot explain the brightness of the Galilean satellites in total umbral eclipses we observed.  
To explain this brightness in total umbral eclipse, \citet{Nakamoto2014} proposes forward-scattered sunlight by haze at the upper region of the Jovian atmosphere 
(higher than several tens mbar, maybe around several mbar region, where are around the top of the stratosphere of Jovian atmosphere).  
This pressure range has been poorly observed by previous spectral observations and in-situ observations by the Galileo probe. 
For example in previous works, \citet{Zhang2013} explicitly stated that 10 mbar is the top of the sensitivity region, and \citet{Banfield1998} stated that the limit is 20 mbar.
Our observation has a potential to resolve the vertical structure of haze well above this pressure level, if this model is true.
Jovian atmosphere of further higher region ($<$1 mbar) was investigated by observations of field star occultation by Jupiter \citep{Raynaud2004, Christou2013}, 
but occultations of field stars by Jupiter are much rarer than the Galilean satellite eclipses.  
Therefore, observations of Galilean satellite eclipsed in Jovian shadow can be a very unique method to investigate the Jovian upper atmosphere, if this hypothesis is true.
The cloud and haze are mainly produced in this pressure range \citep{West1988, Fortney2005}, thus constrain for such pressure range is very important for understanding of Jovian cloud dynamics.

New insights about Jovian atmosphere will be extended to the researches of exo-planets, 
because Jovian atmosphere is the base for atmospheric modeling of exo-planets \citep{Seager2010}.
Especially, since our new technique will provide us the transmission spectrum of Jupiter projected on the Galilean satellites eclipsed in Jovian shadow as screens,
such transmission spectrum of Jupiter will be applied to the modeling of the atmosphere of extrasolar giant planets by transit spectroscopy \citep{Brown2001}.
Ttransit spectra of 11 exo-planets are obtained to date \citep{Swain2014},
and in these spectra, methane \citep{Swain2008} and haze \citep{Pont2008} were first detected in the atmosphere of the hot-Jupiter HD189733b by near-infrared transit spectroscopys.
For example, haze abundance is important to estimate the size of the exo-planet from transit observations \citep{Kok2012}.
Thus, Jovian transmission spectrum, including methane absorption and haze forward-scattering like our data, 
will be a standard of modeling and comparison for characterization of these transit spectra of exo-planets.

\section{Summary}
As seen above, the origin of the brightness of the Galilean satellites in eclipse is still under discussion, but forward-scattered sunlight model \citep{Nakamoto2014} looks more plausible.  
If this is the case, this kind of observations will bring us a new observational method to investigate the composition and time variance of the Jovian atmosphere,
especially abundance of methane and haze, as a function of altitude and position in Jupiter.  
The transmission spectrum of Jupiter by our new method will be applied to the modeling the transit spectrum of exo-planets.

\acknowledgments
\section*{ACKNOWLEDGMENTS}
This research is based on observations made with these instruments; 
the Subaru Telescope which is operated by the National Astronomical Observatory of Japan associated with program S12A-022, S13B-115 and S14A-080; 
the NASA/ESA Hubble Space Telescope, obtained at the Space Telescope Science Institute (STScI), 
which is operated by the Association of Universities for Research in Astronomy, Inc. under NASA contract NAS 5-26555, associated with program \#12980; 
and the Spitzer Space Telescope, which is operated by the Jet Propulsion Laboratory (JPL), California Institute of Technology under a contract with NASA,
associated with program \#80235 and \#90143.  
This work was supported by Japan Society for the Promotion of Science, KAKENHI (\#24111717, \#26800112), and NASA through a grant from STScI and JPL.

\end{document}